\documentclass[letterpaper]{article}
\usepackage[letterpaper, left=1in, right=1in, bottom=1in, top=1in]{geometry}
\usepackage{cmap}
\usepackage{multicol}
\usepackage{cite}
\usepackage{float}
\usepackage{graphicx}
\usepackage[T1]{fontenc}
\usepackage{lmodern}
\usepackage{amsmath}
\usepackage{hyperref}

\title{A hybrid solution for 2-UAV RAN slicing}
\author{Nathan Boyer \\ ENS Paris, LIRMM Montpellier \\ \textit{nathan.boyer@ens.psl.eu}}

\begin{document}

\maketitle

\begin{abstract}

    It's possible to distribute the Internet to users via drones.
    However it is then necessary to place the drones according to the positions of the users.
    Moreover, the 5th Generation (5G) New Radio (NR) technology is designed to accommodate a wide range of applications and industries. The NGNM 5G White Paper \cite{5gwhitepaper} groups these vertical use cases into three categories:
    \begin{enumerate}
        \item enhanced Mobile Broadband (eMBB)
        \item massive Machine Type Communication (mMTC)
        \item Ultra-Reliable Low-latency Communication (URLLC).
    \end{enumerate}

    Partitioning the physical network into multiple virtual networks appears to be the best way to provide a customised service for each application and limit operational costs. This design is well known as \textit{network slicing}.
    Each drone must thus slice its bandwidth between each of the 3 user classes.
    This whole problem (placement + bandwidth) can be defined as an optimization problem, but since it is very hard to solve efficiently, it is almost always addressed by AI in the litterature.
    In my internship, I wanted to prove that viewing the problem as an optimization problem can still be useful, by building an hybrid solution involving on one hand AI and on the other optimization.
    I use it to achieve better results than approaches that use only AI, although at the cost of slightly larger (but still reasonable) computation times.
\end{abstract}

\tableofcontents

\section{Probleme presentation and literature analysis}

According to \cite{main_article}, the concept of network slicing has been explored for some time. With the aim of providing multiple core networks over the same network infrastructure, DECOR and eDECOR \cite{decor} have already been used in legacy LTE networks. 
However, Radio Access Network (RAN) slicing has not been considered in these works.
Since then, the concept of 5G network slicing has evolved to provide better modularisation and flexibility of network functions.
RAN slicing is now of great interest in the literature, as it can support resource isolation between different slices, providing a way to allocate radio resources to the connected User Equipments (UEs) based on the slices they belong to. 
Several RAN slicing approaches have been proposed in recent years. 
Traditional strategies (\hspace{1sp}\cite{RANslicing1}, \cite{orion})  mainly deal with orthogonal resource allocation in time and/or frequency domain, such that the isolation between different services is guaranteed. Their aim in mainly oriented at allocating virtual Resource Blocks (vRBs) to UEs for intra-slice scheduling and to map the allocated vRBs to Physical Resource Blocks (PRBs) at a second stage.

Machine Learning (ML) techniques (\hspace{1sp}\cite{fairness}, \cite{main_article}), mainly deep reinforcement learning, have also been extensively considered, given their capability to find patterns within the huge amount of data that is exchanged in cellular networks.

In \textit{Deep Reinforcement Learning for Combined Coverage and Resource Allocation in UAV-Aided RAN-Slicing}\cite{main_article}, the authors develop a deep reinforcement learning agent that decides the best direction to go for the UAVs over a few 100 steps, and that is the most common approach.
Indeed, in \textit{UAV-Assisted Fair Communication for Mobile Networks: A Multi-Agent Deep Reinforcement Learning Approach}\cite{fairness}, the authors do pratically the same thing but add the dimension of fairness between the users.
Moreover, in \textit{QoS-Aware UAV-BS Deployment Optimization Based on Reinforcement Learning}\cite{deep-qn}, the authors show that using deep-Q-network, it is possible to achieve optimal performance when placing only 1 base station.
Finally, in \textit{Reinforcement Learning Aided UAV Base Station Location Optimization for Rate Maximization}\cite{rate_max}, the authors use again deep reinforcement learning to resolve the placement of 3 UAVs and find that it achieves around 70\% of the optimal performance, and that it is better than a k-means algorithm when the measure of imperfection is higher.

\;

My internship aims at developping the following points on a 2-UAVs-aided RAN slicing:

\begin{enumerate}
    \item Improving the optimization algorithm to make it able to generate big amount of data in reasonable time
    \item Developping an AI agent that learns from the optimal optimization algorithm with supervised learning
    \item Building an hybrid agent from this AI agent that combines AI and a classical optimization algorithm to achieve better performance
    \item Realizing an honest comparison between the different methods both in terms of performance achieved and in terms of computation time
\end{enumerate}

\section{Problem definition and solution via mathematical optimization}

\label{optimization}

In \cite{main_article} it is showed how to model the problem considered with a mathematical model.

First, we need to define the equation that calculates how well a user will receive a connection from a base station, depending on where it is.

We have the following equations for the SINRs of the drones, which is a factor that represents how much data is actually received by the user in relation to the data sent to the user by the base station:

$\operatorname{SINR}_{i, j}^t=\frac{p c \mu\left(y_j, u_i^t\right)\left(\left(\left\|y_j-u_i^t\right\|\right)^2+\left(h\right)^2\right)^{-\alpha / 2}}{\sum\limits_{k \in \mathcal{U} \backslash i} p c \mu\left(y_j, u_k^t\right)\left(\left(\left\|y_j-u_k^t\right\|\right)^2+\left(h\right)^2\right)^{-\alpha / 2}+\sigma^2}$

To evaluate a configuration, we will therefore proceed as follows:

\begin{itemize}
    \item First, each user must be associated with the nearest base station.
    \item You then divide the bandwidth of a base station dedicated to a given class between all the users of that class associated with that base station.
    \item You then calculate the SINR for each user and decide whether the user is satisfied or not.
    \item The percentage of satisfied users is what you want to maximise.
\end{itemize}

\subsection{Optimization problem}

This problem can be resolved as an optimization problem, I will show you how to resolve it for 2 base stations.

Let's first define $u$ a matrix of dimensions $N*N$, with $N$ being the number of positions possible for a base station.
$u_{i,j}$ is a binary variable that is equal to one if and only if the first base station is in position i and the second is position j.

$bw$ is a $3*2$ matrix that represents the bandwidth each base station allocates to each user class.

$\delta$ is a vector that represents for every whether it is satisfied or not.

The function to optimize is then: $\max_{u,bw}\sum_{g\in\mathcal{G}}\delta_g$

Under the following constraints:

\begin{subequations}
    \allowdisplaybreaks
    \begin{align}
        \sum_{b \in |\mathcal{U}|} \sum_{i, j \in \mathcal{U}} u_{i, j} \times bw_{\text{slice}(g), b} \times \text{bps}_g \frac{100}{G_{\text{conn}}(\text{slice}(g),b)} \nonumber
        \geq th_g \times \delta_g \hspace{1cm} &\forall g \in \mathcal{G} \\[1em]
        bw_{em,b} + bw_{ur,b} + bw_{mm,b} = 1 \hspace{1cm} &\forall b \in |\mathcal{U}| \\[1em]
        bw_{em,b} \geq 0; bw_{ur,b} \geq 0; bw_{mm,b} \geq 0 \hspace{1cm} &\forall b \in |\mathcal{U}| \\[1em]
        \sum_{i,j \in \mathcal{U}}u_{i,j} = 1 &\\[1em]
        \delta_g \in \{0, 1\} \hspace{1cm} &\forall g \in \mathcal{G} \\[1em] 
        u_{i,j} \in \{0, 1\} \hspace{1cm} &\forall i,j \in \{0, 1, ... , |\mathcal{U}-1|\}
    \end{align}\label{constr}
    \end{subequations}

The resolution of this optimization problem can be implemented in Python with the help of a module such as pulp.

\subsection{Instances generation}

This internship is mainly based on this article: \textit{Deep Reinforcement Learning for Combined Coverage and Resource Allocation in UAV-Aided RAN-Slicing}\cite{main_article}.
As such, instances of the problem are generated in a very similar way as they are in this article, and are as follows:

\begin{table}[H]
    \centering
    \begin{tabular}{|c|c|}
    \hline
    Parameter & Value \\
    \hline
        Number of users $n$ & 50 (later 25 - 100)\\
        Number of clusters $c$ & 2 (later 0 - 5) \\
        UAV transmit power $p$ &  1 W\\
        pathloss exponend $\alpha$ & 2.1 \cite{Channel_Model}\\
        UAV half-power beamwidth $\eta$ & 30 deg. \cite{Channel_Model}\\
        Near-field pathloss $c$ & -38.4 dB \cite{Channel_Model}\\
        Noise power $\sigma^2$ & $8 \cdot 10^{-13}$ W\\
        Bandwidth $B$ & 20 MHz\\
        Users density, $|\mathcal{G}|/\text{km}^2$ & 100\\
        UAVs number, $|\mathcal{U}|$ & 2\\
        UAVs height $h$ & 50 m\\
        UAVs density $|\mathcal{U}|/\text{km}^2$ & 8\\
        eMBB users demand $\text{dem}_\text{eMBB}$ & 5 Mbps \cite{demandsvalues}\\
        URLLC users demand $\text{dem}_\text{URLLC}$ & 10 Mbps \cite{demandsvalues}\\
        mMTC users demand $\text{dem}_\text{mMTC}$ & 0.5 Mbps \cite{demandsvalues}\\
        $p_{\text{eMBB}}$, $p_{\text{URLLC}}$, $p_{\text{mMTC}}$ & 20\%, 10\%, 70\% \\
    \hline
    \end{tabular}
    \caption{Environment Parameters}
    \label{tab:envparams}
\end{table}

\subsection{Improvement of the optimization algorithm}

In this section we present an effective preprecessing of the data that can be used to obtain a good
tradeoff between quality of the solution and computing time.

First, only positions in the convex envelope of the users can actually be candidates for optimal base station placements.
This allows us to significantly reduce the size of the matrix u.

It is also possible to go further, but not without losing precision.
In fact, we can divide the users into k clusters and take the union of this convex hull of these clusters.

The transformation carried out can be visualised as follows:

\begin{figure}[H]
    \centering
    \includegraphics[width=0.5\textwidth]{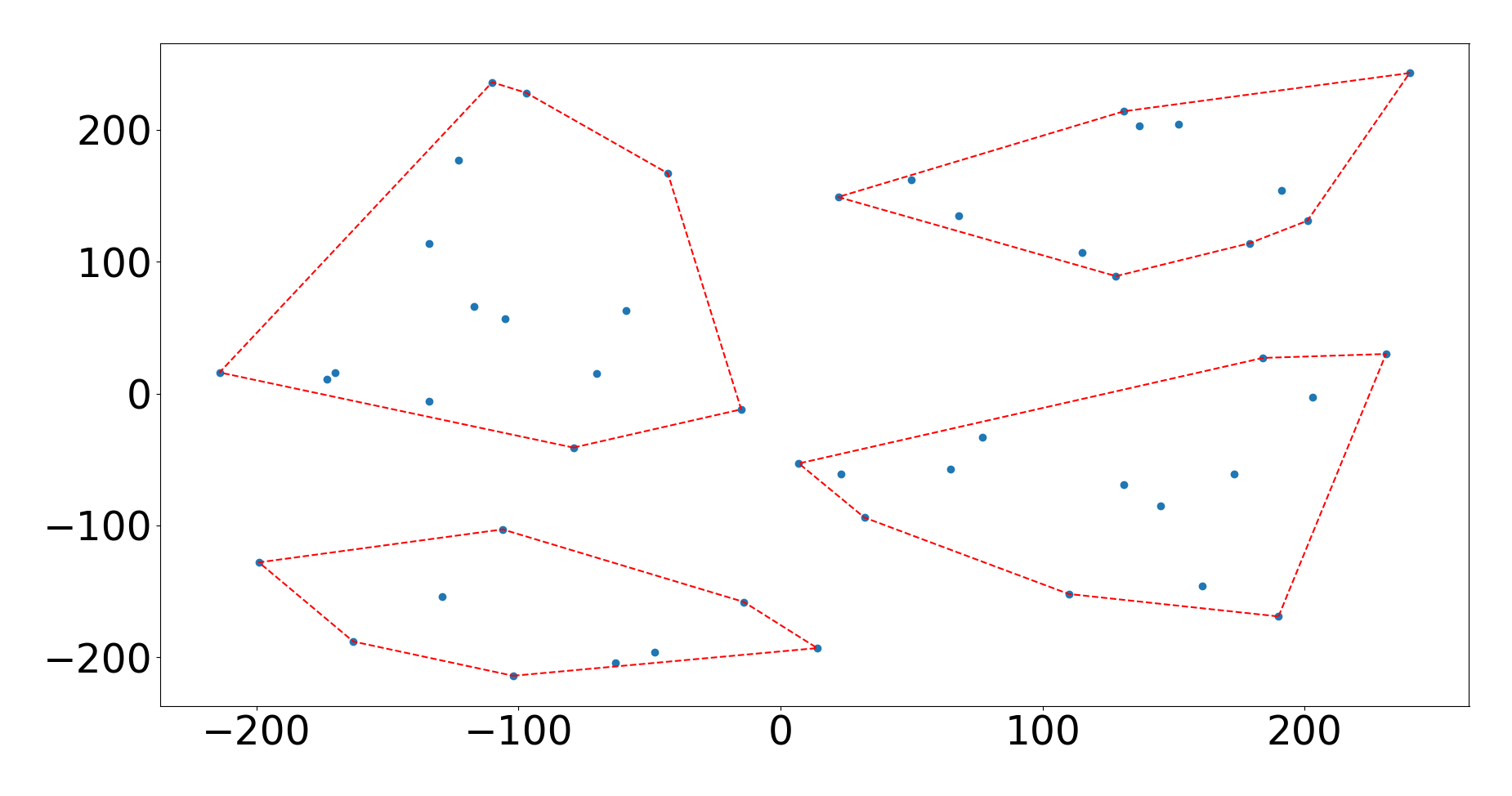}
    \caption{Representation of the convex enveloppe of the 4 clusters}
\end{figure}

The graph of performance and time as a function of the number of clusters is shown below for both random positions and positions generated with 2 clusters.

\begin{figure}[H]
    \centering
    \begin{minipage}[b]{0.45\textwidth}
        \centering
        \includegraphics[width=\textwidth]{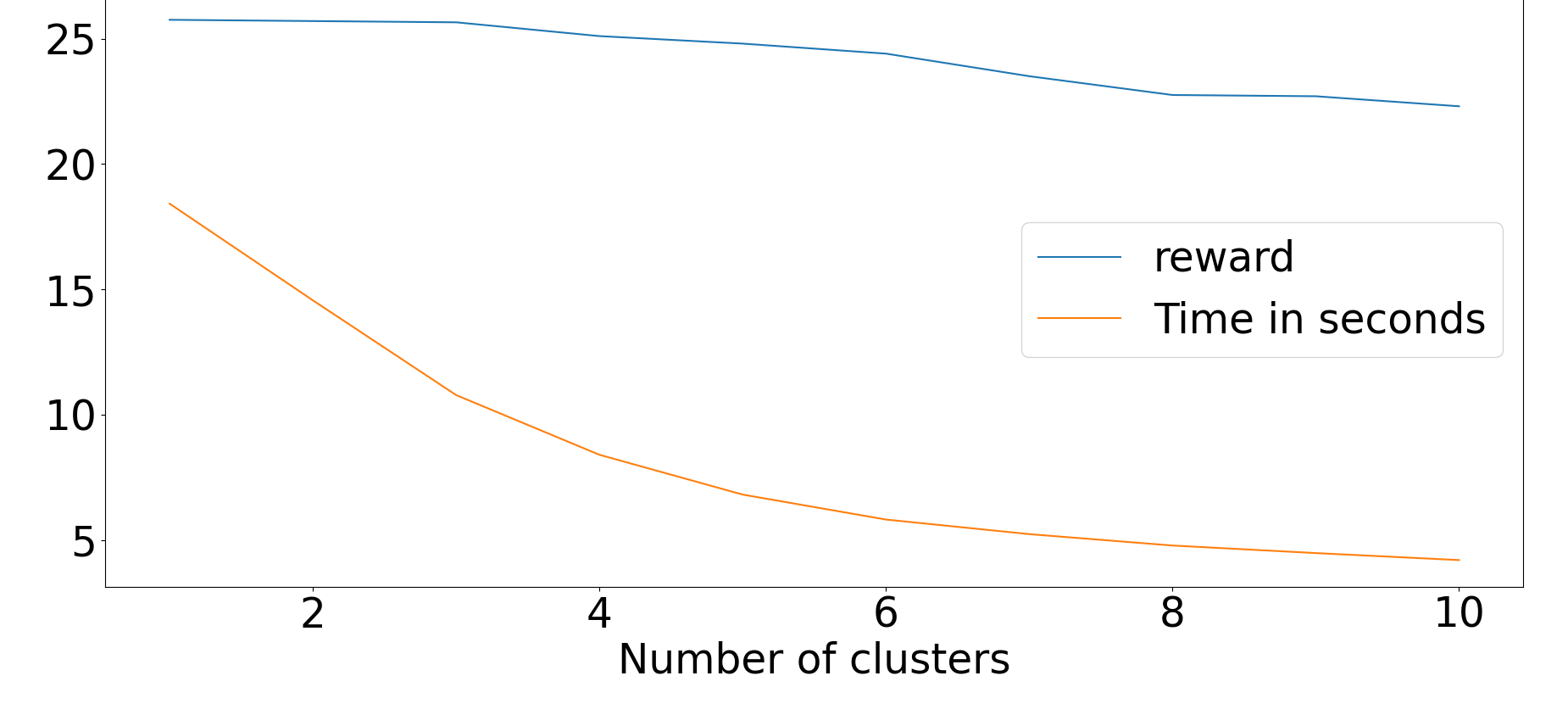}
        \caption{Performance when users are random}
    \end{minipage}
    \hspace{0.05\textwidth}
    \begin{minipage}[b]{0.45\textwidth}
        \centering
        \includegraphics[width=\textwidth]{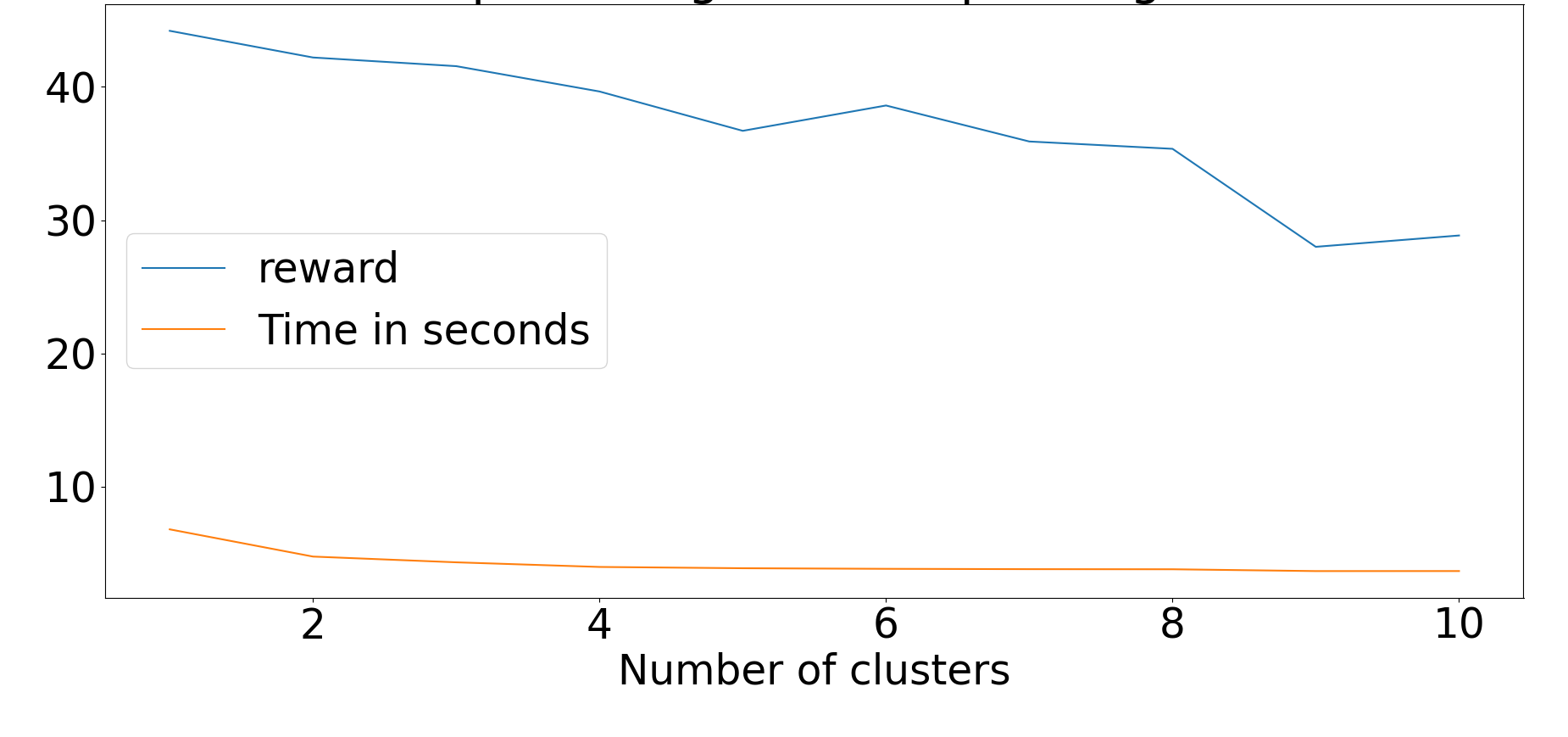}
        \caption{Performance when users are in 2 clusters}
    \end{minipage}
\end{figure}

However, since the computation times are pretty reasonable for 2 base stations, I decided to go with only 1 cluster.

Finally, multithreading can be used to further accelerate the computations.

\section{Machine learning to solve the problem}

However, the model presented in the previous sectioncan still require a signficant amount of time to besolved.
This can be problematic, especially if the user is moving and the optimal position needs to be recalculated frequently.

A machine learning solution is therefore considered.
In fact, solving an instance of the problem would only require a forward pass of the neural network, which is negligible in time compared to solving an instance with the optimal agent.

\subsection{Reinforcement learning}
\label{reinforcement}

A reinforcement learning solution is designed and implemented, in article\;\cite{main_article}

The authors tried to solve the problem I just presented and, in short, managed to get an average reward of 72-73\% of users satisfied.

However, there are several differences between their approach and mine:

\begin{enumerate}
    \item First, they move the base stations step by step and computes the mean reward over 100 movements.
          However, since the base stations spend most of their times idling (because they have already reached their best possible position),
          I took the bias of putting the base stations directly at the computed best position, and then computing the reward only once (which also allows to test the different agents over more instances in the same amount of time).
    \item Secondly, what is shown in their article is the average reward during the reinforcement process.
        In other words, the neural network is only tested on the instances that it has just learnt from.
        What I have done instead is to have a separate test set that I use to evaluate the different agents.
\end{enumerate}

\subsection{Supervised learning from the optimal solution}

In this section we present how to use supervised learning to train an agentthat leanes from the optimal solutionsprovided by the mathematical model
presented in \hyperref[optimization]{section 2}.
By using an optimal agent, the neural network can learn to replicate the optimal solutions.
Developing such an AI was the first topic of my internship.

The neural network is therefore divided into 2 neural networks:

\begin{enumerate}
    \item A neural network that takes the positions of the users as input and learns to output the best positions for the 2 base stations.
    \item A neural network that takes as input the positions of the users and the positions of the base stations and learns to output the best allocation of bandwidth between the 3 classes of users.
\end{enumerate}

As an aside, each user's type is represented by 3 one-hots, one for each class.
More details on how the neural networks are built are available in the \hyperref[appendix]{appendix}

To conclude my results, here is the average error for the neural network that learns to output the positions:
As well as the average error for the neural network that learns to output the bandwidths:

\begin{figure}[H]
    \centering
    \begin{minipage}[b]{0.45\textwidth}
        \centering
        \includegraphics[width=\textwidth]{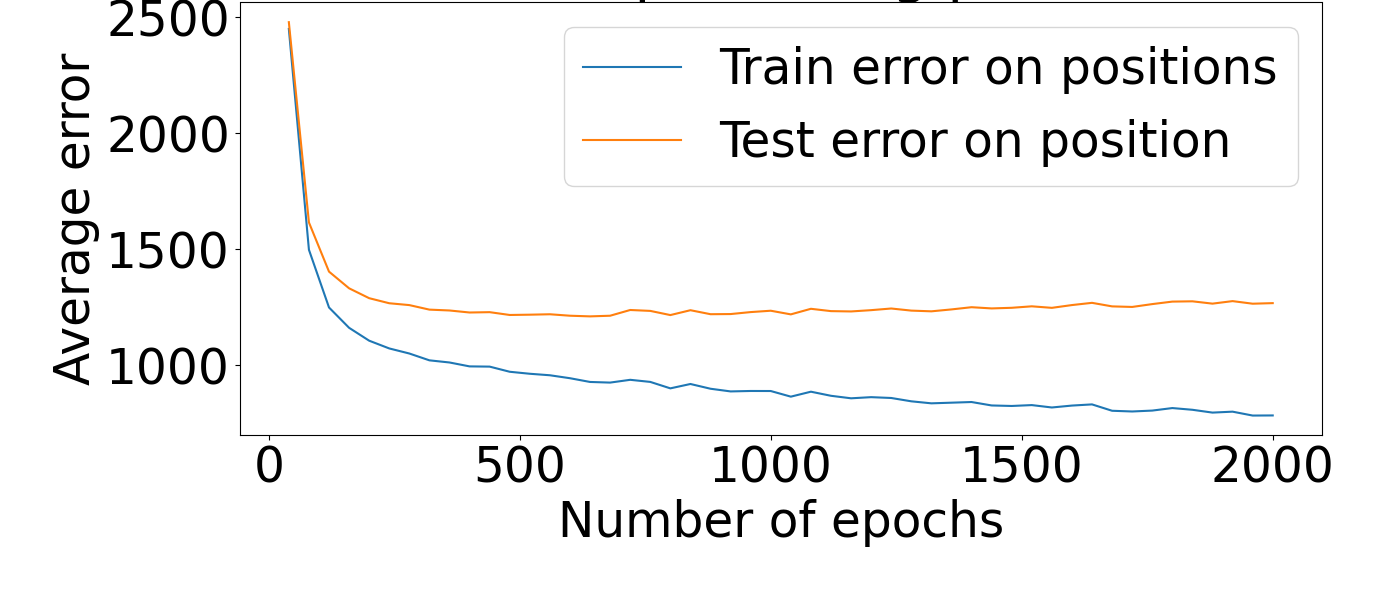}
        \caption{Position error}
        \label{fig:image5}
    \end{minipage}
    \hspace{0.05\textwidth}
    \begin{minipage}[b]{0.45\textwidth}
        \centering
        \includegraphics[width=\textwidth]{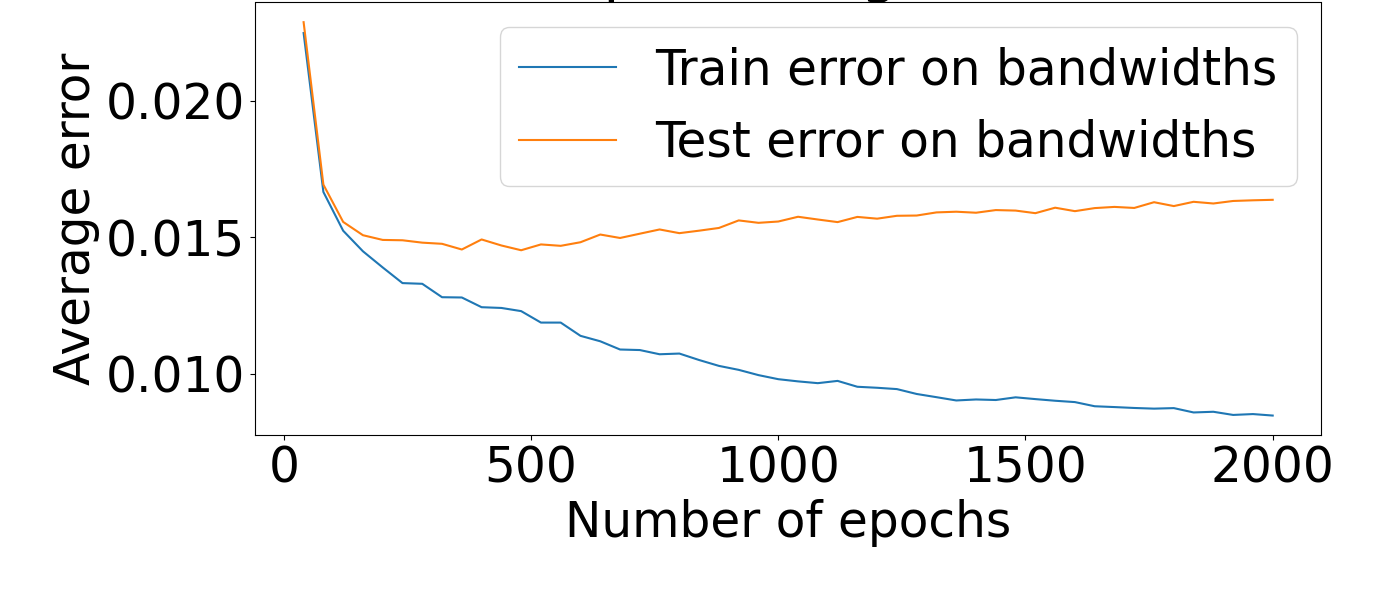}
        \caption{Bandwidth error}
        \label{fig:image6}
    \end{minipage}
\end{figure}

\;

With these neural networks, we can test them on instances and evaluate the performance they give.

(We plot here the normalized average user coverage, i.e.\, the average user coverage divided by the average user coverage of the optimal solution)

\begin{figure}[H]
    \centering
    \includegraphics[width=0.5\textwidth]{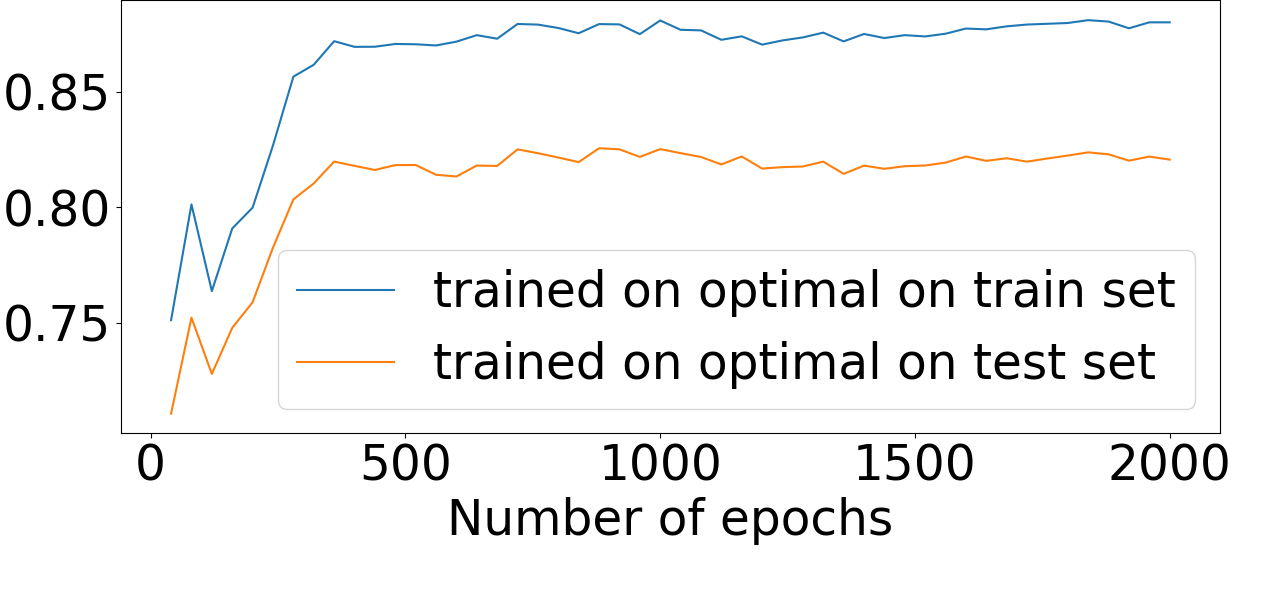}
    \caption{Normalized average user coverage for AI agent}
\end{figure}

\subsection{Learn only what's necessary}

From \hyperref[fig:image6]{figure 5} it is possible to see that the test error on the bandwidths stops decreasing rapidly.
That's why I decided to show the performances when the positions are optimally decided (i.e. decided by the optimization algorithm), but the bandwidths are decided by the AI.

\begin{figure}[H]
    \centering
    \includegraphics[width=0.5\textwidth]{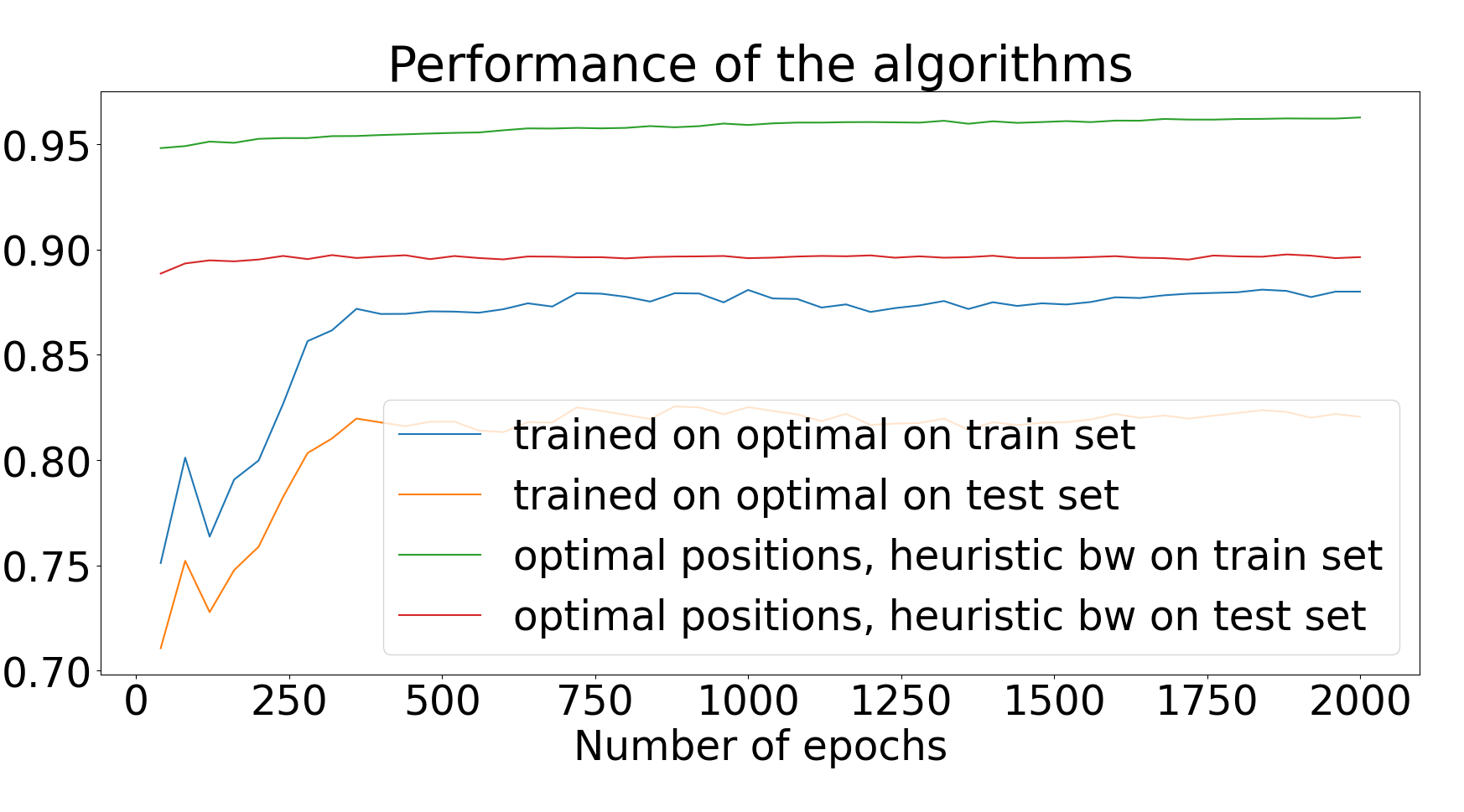}
    \caption{Performance with optimal positions and AI bandwidths}
\end{figure}

As we can see, the bandwidth part seems to be the problem, because even with perfect positions, the AI agent's solution is still
is still far from perfect.

So I implemented an agent that decides the positions with a neural network, but decides the bandwidth optimally,
which is quite fast once the positions are fixed.

Then we have the following results:

\begin{figure}[H]
    \centering
    \includegraphics[width=0.5\textwidth]{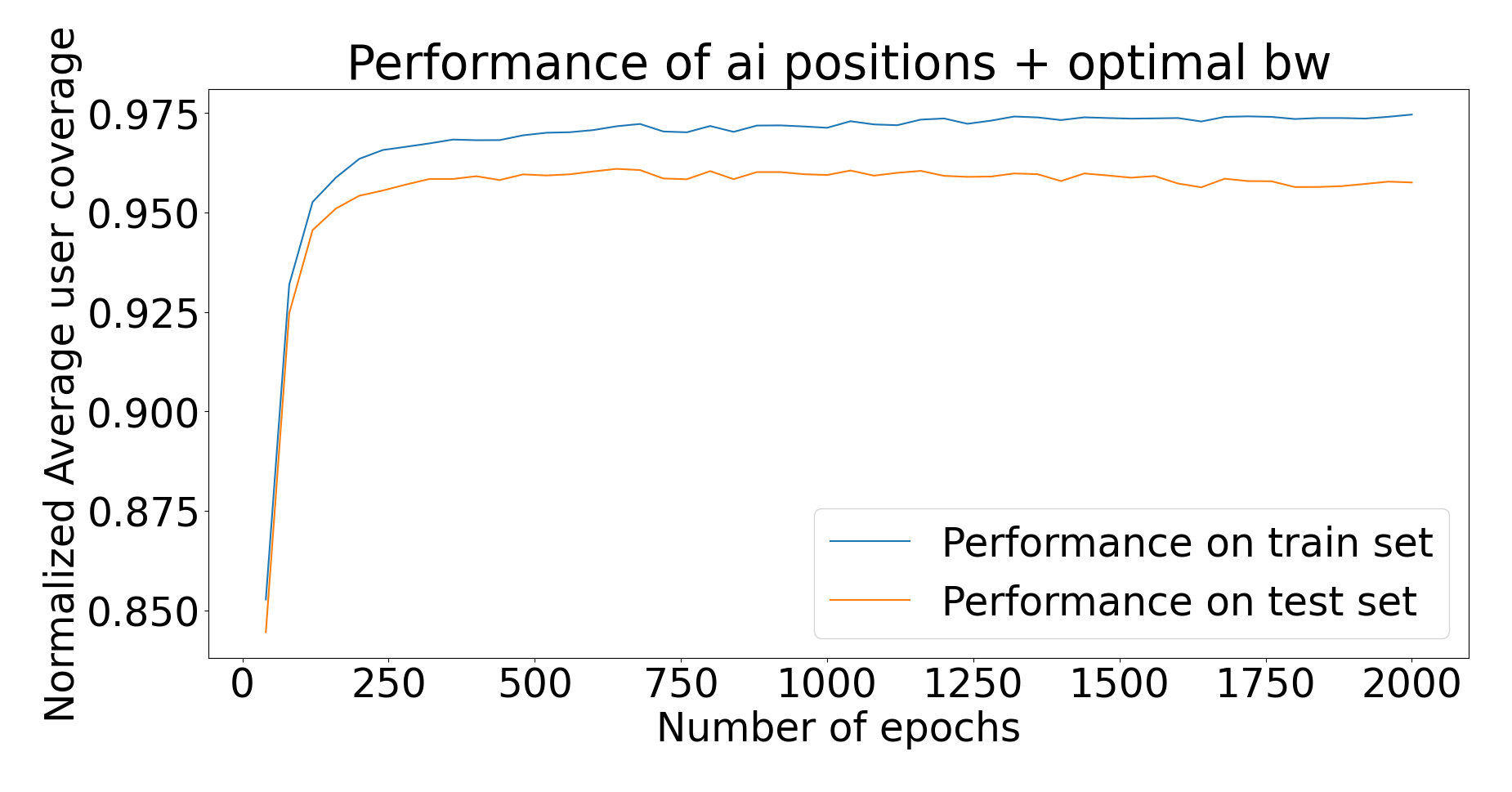}
    \caption{Performance with AI positions and optimal bandwidths}
\end{figure}

The results here are pretty convincing, but I will do a deep and honest analysis of the different methods in terms
of average reward but also in terms of average computation time.

\subsection{Generalizing to a variable number of users}

In the previous sections, I have only solved instances with a fixed number of users (50).

The first solution we came up with was to use graph neural networks.
Graph Neural Networks (GNNs) are neural networks that consist of a variable (i.e.\, not fixed) number
of nodes and edges and perform some aggregation functions (such as mean, min, or max) on them to get a final answer.

The graph given to the neural network was constructed using a k-neighbourg algorithm.
I made a graph showing the final test error on positions as a function of the k in the k-neighbourg algorithm:

\begin{figure}[H]
    \centering
    \includegraphics[width=0.5\textwidth]{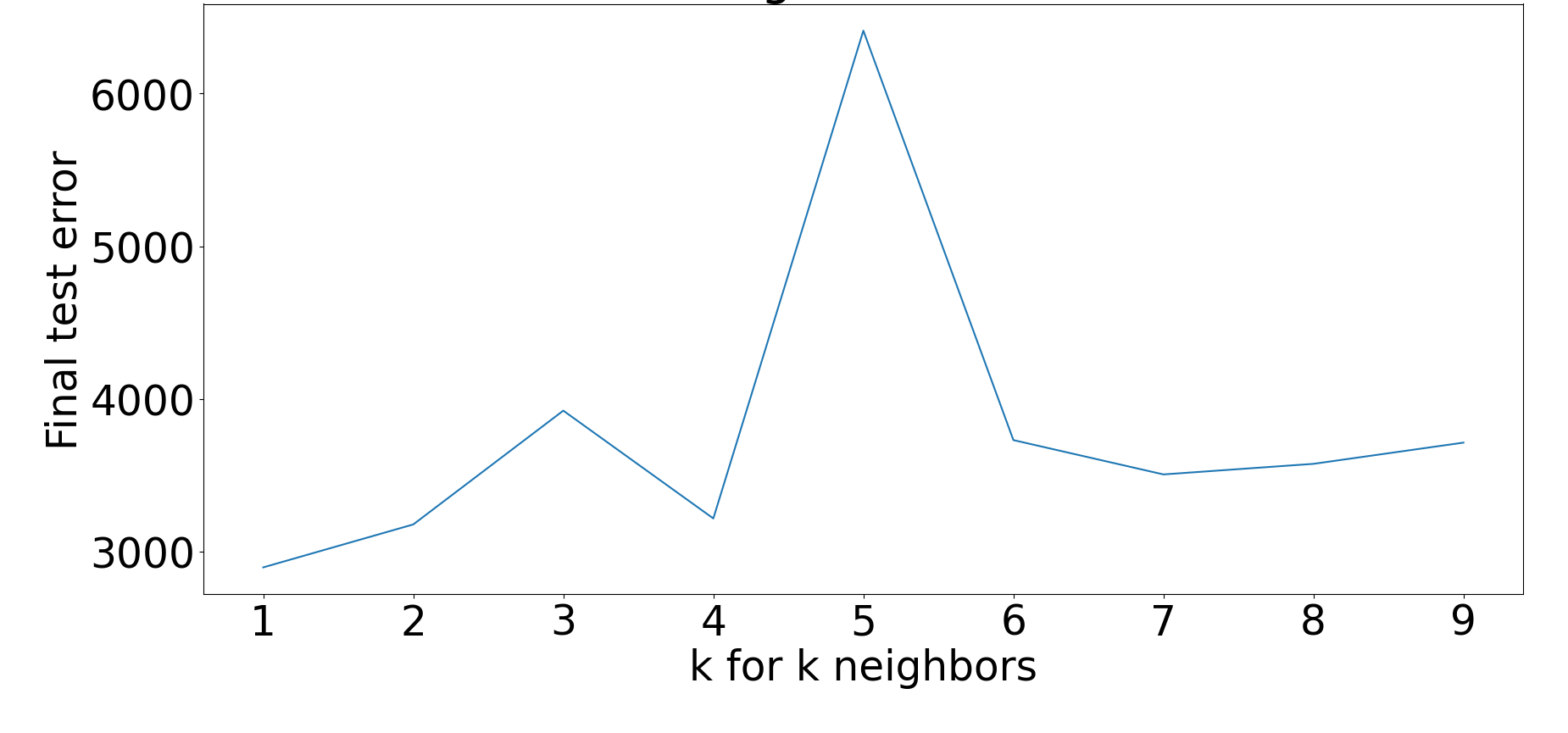}
    \caption{Position error for graph neural network in function of k}
\end{figure}

Unfortunately, regarless of the choice for k, the the error obtained is always significantly higher than the one obtained with a model for a fixed
number of users

So I decided on a simpler but less scalable approach: I simply add a one-hot to each user, indicating whether the user exists or not.
Thus, the neural network can be used to resolve instances with a variable number of users, as long as the number in question is less than a certain maximum.

The accuracy with a variable number of users from 25 to 100 is very similar to the accuracy with a fixed number of users, as you can see in the following graphs:

\begin{figure}[H]
    \centering
    \begin{minipage}[b]{0.45\textwidth}
        \centering
        \includegraphics[width=\textwidth]{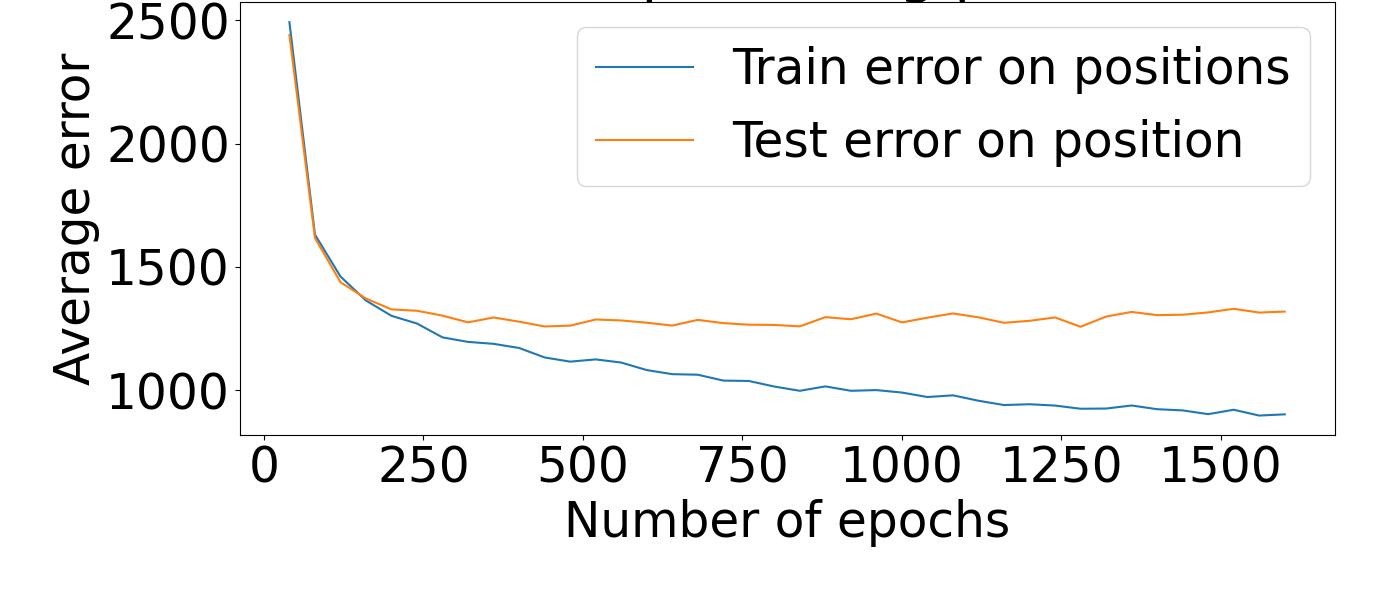}
        \caption{Position error}
    \end{minipage}
    \hspace{0.05\textwidth}
    \begin{minipage}[b]{0.45\textwidth}
        \centering
        \includegraphics[width=\textwidth]{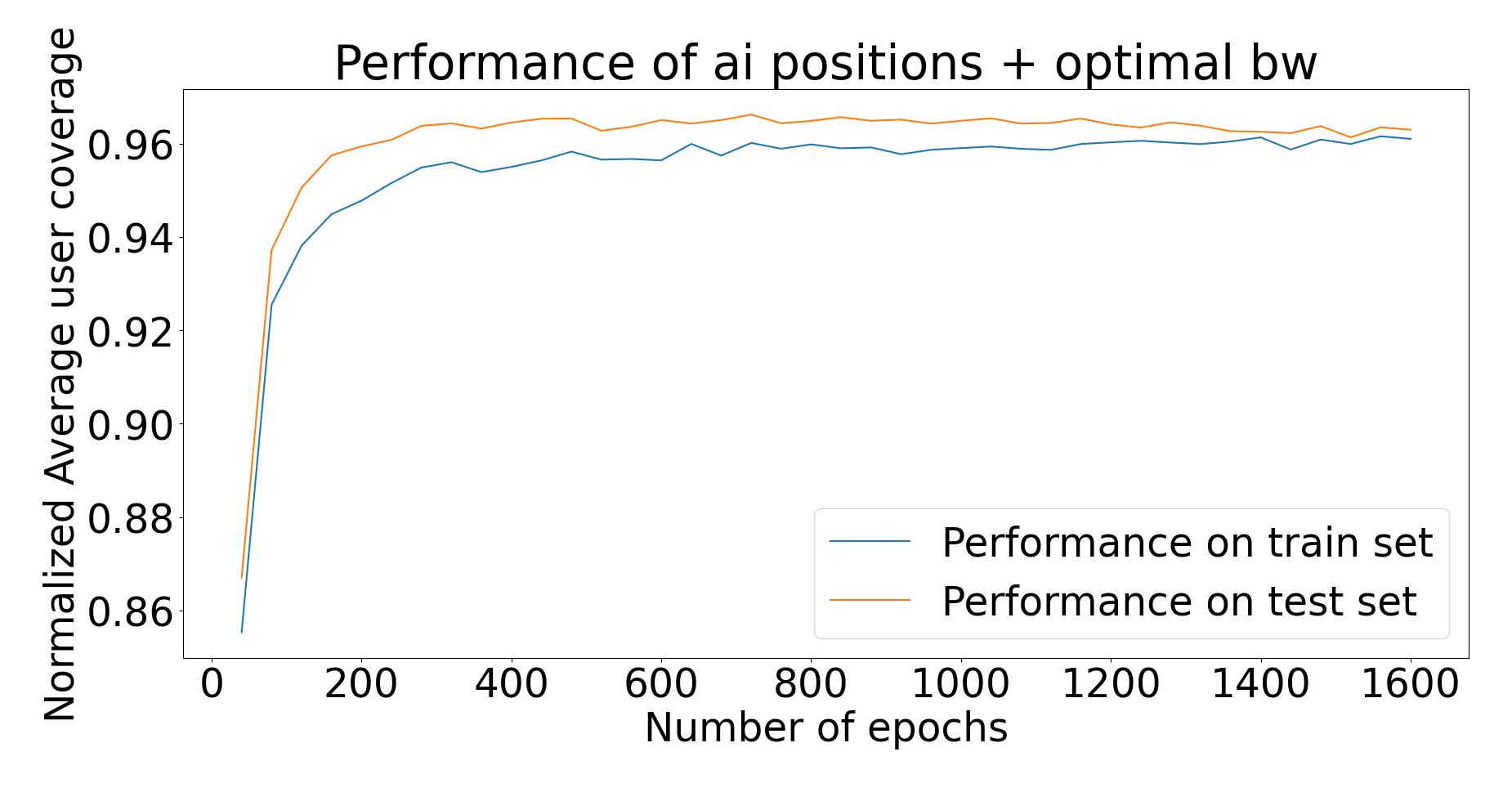}
        \caption{Average normalized reward}
    \end{minipage}
\end{figure}

\subsection{Comparison of the different methods}

In this subsection I will compare the 3 agents we have seen in this report: the optimal agent, the full AI agent, and the hybrid agent.
The 2 criteria will be first average reward and second time computation.
Also, so far I have always generated users in 2 clusters, which is the simplest case (and the one used in the article my internship is based on \,\cite{main_article}).
However, I will now evaluate the different agents on a number of clusters ranging from 0 (meaning random) to 4.

We first evaluate the average reward, i.e. the average percentage of users satisfied.

\begin{center}
    \begin{tabular}{|c|c|c|c|c|c|}
    \hline 
     Type of agent & random & 1 cluster & 2 clusters & 3 clusters & 4 clusters \\ 
     \hline
     Optimal & 50.25\% & 57.80\% & 87.28\% & 76.18\% & 57.77\% \\  
     \hline
     Full-ai & 32.81\% & 46.41\% & 77.01\% & 57.08\% & 49.73\%  \\
     \hline
     Hybrid &  36.82\% & 55.87\% & 83.62\% & 64.60\% & 55.84\%  \\
     \hline
    \end{tabular}
\end{center}

While in some cases, the results are a bit lacking (namely in the random and 3-clusters cases), the results are generally pretty close to the optimal solution.
On the other hand, the hybrid option is always better than the full AI.

Moreover, in the case that is the most similar to the one the situation in the article I tried to improve upon\,\cite{main_article}, while they had a normalized user coverage
of around 87\%, we have one of 96\% ($=\frac{83.62}{87.28}$). And even if the situations are not exactly the same (\hyperref[reinforcement]{details}), the changements should normally make it harder for my agent,
since evaluating on the optimal positions makes it harder to be optimal rather than just having to move towards it,
and having a separate test set punished overfitting harder.

To obtain a fair comparison of the methodologies proposed we need to consider also the computiong time needed

The time computations for the optimal agent depend of number of clusters because of the optimizations made.

\begin{table}[!h]
\begin{center}
    \caption{Computation times for the optimal agent}
    \begin{tabular}{|c|c|c|c|c|c|}
    \hline 
       & random & 1 cluster & 2 clusters & 3 clusters & 4 clusters \\ 
     \hline
     Average computation times & 8.48s & 1.49s & 1.79s & 2.36s & 2.86s \\  
     \hline
    \end{tabular}
\end{center}
\end{table}

However, it is stable for the other 2 agents, with an average computation time of about 8-9 ms for the full-AI agent and an average computation time of about 25-30 ms for the hybrid agent,
and an average computation time of about 25-30 ms for the hybrid agent.

Thus, even if the hybrid agent can be seen as a middle ground between the short computation times but low rewards of the full-AI agent and the high rewards but long computation times of the full-AI agent, the hybrid agent still performs well,
and the high rewards but long computation times of the optimal agent, I think the balance is in its favor,
with only slightly longer computation times than the full-AI agent and slightly worse rewards than the optimal agent.

\section{Areas for improvement}

First, the AI agents were only trained with a data set of 9000 instances, which could be increased if needed,
and the rewards would be better.

Second, a relatively new idea in solving optimization problems with AI is to use the distance to the optimal solution.
(but in terms of reward, not in terms of answer as we did) as the reward for the neural network, which could definitely be implemented for this problem,
and might give higher performance.

Thirdly, as seen in this article\cite{main_article}, the reward the optimal agent has is stongly tied to the precision of the grid used.
I used a grid of $10 \times 10$ in my internship but by increasing the precision, the rewards would be better.

Finally, if you wanted to extend the method presented in this article to a number of drones of 3 or more, you would have to
rewrite the optimal agent completely, as it scales very poorly.
For example, it currently requires 37.1 GB of ram to declare the $u$ matrix for 4 drones.

\section{Conclusion}

In my internship, I developed 2 RL agents for a 2-UAV assisted 5GNR network slicing scenario, with the goal of optimizing the SLA satisfaction of the deployed users.

The goal was to learn from the optimal agent, with the idea that it is not really the movements of the UAVs that are important,
but their final destinations.

I demonstrated how a hybrid approach can outperform a full-AI approach, which is commonly used in the literature.

Finally, this method showed great adaptability, performing well with many different distributions of users.

The promising results obtained during my internship can encourage us to investigate towards improving the performance,
and try to scale this method to situations with more than 2 UAVs.

\bibliographystyle{plain}
\bibliography{bibliography}

\begin{thebibliography}{10}

\bibitem{decor}
TR~23.711 3GPP.
\newblock Enhancements of dedicated core networks selection mechanism (release
  14), September 2016.

\bibitem{orion}
Xenofon Foukas, Mahesh Marina, and Kimon Kontovasilis.
\newblock Orion: Ran slicing for a flexible and cost-effective multi-service
  mobile network architecture.
\newblock In {\em unknown}, pages 127--140, 10 2017.

\bibitem{Channel_Model}
Boris Galkin, Babatunji Omoniwa, and Ivana Dusparic.
\newblock Multi-agent deep reinforcement learning for optimising energy
  efficiency of fixed-wing uav cellular access points, 2021.

\bibitem{rate_max}
Sudheesh~Puthenveettil Gopi and Maurizio Magarini.
\newblock Reinforcement learning aided uav base station location optimization
  for rate maximization.
\newblock {\em electronics}, 2021.

\bibitem{RANslicing1}
Adlen Ksentini and Navid Nikaein.
\newblock Toward enforcing network slicing on ran: Flexibility and resources
  abstraction.
\newblock {\em IEEE Communications Magazine}, 55:102--108, 06 2017.

\bibitem{deep-qn}
Hyungje Lee, Chahyeon Eom, and Chungyong Lee.
\newblock Qos-aware uav-bs deployment optimization based on reinforcement
  learning.
\newblock {\em International Conference on Electronics, Information, and
  Communication}, 2023.

\bibitem{demandsvalues}
Rongpeng Li, Zhifeng Zhao, Qi~Sun, Chih-Lin I, Chenyang Yang, Xianfu Chen,
  Minjian Zhao, and Honggang Zhang.
\newblock Deep reinforcement learning for resource management in network
  slicing.
\newblock {\em IEEE Access}, 6:74429--74441, 2018.

\bibitem{main_article}
Emiliano~Traversi Lorenzo~Bellone, Boris~Galkin and Enrico Natalizio.
\newblock Deep reinforcement learning for combined coverage and resource
  allocation in uav-aided ran-slicing.
\newblock {\em IEEE}, 2022.

\bibitem{5gwhitepaper}
{NGMN}.
\newblock {NGMN} {5G} white paper.
\newblock {\em {NGMN} Alliance}, page 125, 2015.

\bibitem{fairness}
Yi~Zhou, Zhanqi Jin, Huaguang Shi, Zhangyun Wang, Ning Lu, and Fuqiang Liu.
\newblock Uav-assisted fair communication for mobile networks: A multi-agent
  deep reinforcement learning approach.
\newblock {\em remote sensing}, 2022.

\end{thebibliography}

\appendix

\section{Appendix detailling how the neural networks are built}
\label{appendix}

For the position agent, each user is represented as:
\begin{enumerate}
    \item Its x and y axis
    \item A one-hot for each of the possible type
    \item A one-hot that indicates whether the user actually exists (if the user does not exist, the positions are set to random and the one-hots above are set to 0).
\end{enumerate}

The number of users is easily adjustable (I set it to 100 in the tests).
The neural network consists of 2 convolutional layers followed by 2 linear layers.
The hyperparameters are the following: learning rate of $5e^{-5}$ and weight decay of $1e^{-5}$.

\;

For the bandwidth agent, each user is represented as before, but an entry is added for the position of each of the UAVs.
The neural network then consists of a convolutional layer for the drones, a convolutional layer for the users,
then the 2 results are concatenated and passed to 2 linear layers.
The hyperparameters are the following: learning rate of $2e^{-4}$ and weight decay of $1e^{-6}$.

\end{document}